\documentclass[%
 reprint,
 nofootinbib,
 nobibnotes,
 longbibliography,
 amsmath,amssymb,
 aps,
 pra,
floatfix,
]{revtex4-2}

\usepackage{graphicx}
\usepackage{dcolumn}
\usepackage{bm}
\usepackage{textcomp,gensymb} 

\begin{document}

\preprint{APS/123-QED}

\title{Measurements of the Kr (e,~2e) differential cross section in the perpendicular plane, from 2~eV to 120~eV above the ionization threshold}

\author{Andrew James Murray}
\author{Joshua Rogers}
 \email{andrew.murray@manchester.ac.uk, joshua.rogers@manchester.ac.uk}
\affiliation{
 Photon Science Institute, Department of Physics and Astronomy\\
 University of Manchester, 
 Manchester, M13 9PL, UK
}

\date{\today}

\begin{abstract}
New (e,~2e) differential cross section measurements from krypton are presented in the perpendicular plane, where the incident electron beam is orthogonal to the scattered and ejected electrons that map out a detection plane. New data were obtained at incident energies from 30 eV to 120 eV above the ionization potential (IP), the experiment being configured to detect scattered and ejected electrons with equal energy. The results are compared to previous measurements from 2 eV to 50 eV above the IP and to calculations from different models in this energy range. The new experiments confirm the results from previous measurements. The results are also compared to recent data for argon acquired under the same kinematic conditions, to highlight similarities and differences that are observed. 

\end{abstract}

\maketitle

\section{Introduction}

\begin{figure}[h]
\includegraphics[width=8.6cm]{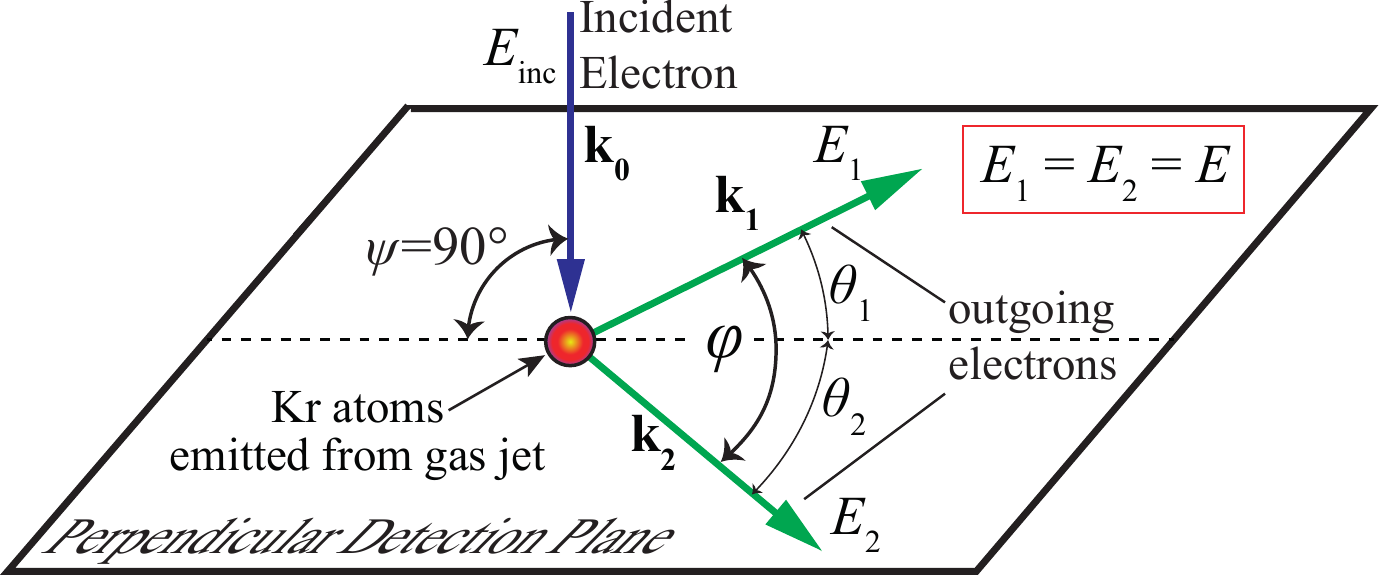}
\caption{\label{fig:e2ePerpendicularPlane} The perpendicular plane scattering geometry.}
\end{figure}

In a recent paper~\cite{Argon2022} the electron-impact ionization differential cross-sections (DCS) from an argon target were detailed, where the scattered and ejected electrons from the interaction were detected in the plane perpendicular to the incident electron momentum \textbf{k}$\bm{_0}$, as shown in Fig.~\ref{fig:e2ePerpendicularPlane}. The outgoing electrons with momenta \textbf{k}$\bm{_1}$ and \textbf{k}$\bm{_2}$ were detected in coincidence with equal energy, using hemispherical electron analyzers that swept around the detection plane. In this plane only the mutual angle $\phi$ has relevance. 

The results from this argon study confirmed previous experimental data~\cite{nixon2010low}, which found a significant disagreement with the calculations carried out by Whelan and co-workers~\cite{miller2015energy}. This was particularly noticeable around 50~eV above the ionization potential (IP). The models included non-relativistic distorted wave Born approximation (DWBA) calculations including and neglecting post-collisional interactions ($\pm$PCI), and a plane wave approximation (PWA). PCI was included using a Gamow factor~\cite{Gamow1928} and also by including a Ward-Macek factor~\cite{WardMacek1994}. Both factors provide an approximation to the effects of PCI and their inclusion was seen to overcompensate the effects of electron-electron repulsion under these kinematic conditions.  The Ar experimental data were further extended to an incident energy of 200~eV above the ionization potential in~\cite{Argon2022}, allowing future tests of different models.

A comparison with low energy experimental data for Kr as presented in~\cite{nixon2010low} was also carried out in~\cite{miller2015energy}. When compared to the Ar studies, the DWBA-PCI calculation was found to be in better agreement for Kr, particularly at an incident energy 50~eV above the IP. This contrast between the calculations for these different targets hence suggested that an extended survey of the cross section for krypton should also be carried out, both to check the data in~\cite{nixon2010low} and to extend the measurement range to higher energies. 

The results of these experiments are presented in this paper, together with a comparison to previous data from~\cite{nixon2010low} and with the calculations in~\cite{miller2015energy}. The energy range of the original study was 2~eV to 50~eV above the IP. The new measurements detailed here extend this range to 120~eV above the IP. This range could not be extended further due to the very low coincidence count rates encountered in this geometry. Selected measurements of the DCS from~\cite{nixon2010low} at incident energies 30, 40 and 50~eV above the IP were repeated and compared to the new measurements. Agreement was found between the two data sets at these energies. 

In all of these studies the DCS are not measured on an absolute scale. The mutual angle $\phi$ = 180$\degree$ is therefore selected as a normalization point. This is a common angle for all incident beam angles $\psi$ since the outgoing electrons are detected in a symmetric geometry (i.e. $\theta_{1}$~= $\theta_{2}$, see Fig.~\ref{fig:e2ePerpendicularPlane}). Normalizing the data at this point allows the experimental results and calculations to then be directly compared over a wide range of geometries, as discussed in~\cite{murray1993evolution,murray1992_PRL}.

In the present studies the energies of the detected electrons were chosen to be equal, so that $E_1$ = $E_2$ = $E$.  The incident electron energy was hence set to be $E_{inc}$ = 2$E$ + IP. An (e,~2e) DCS was then determined that depends on both the energy and mutual angle $\phi$.    

In the perpendicular plane geometry the ionization reaction is highly sensitive to multiple order scattering processes~\cite{al2009atomic,ren2010tracing} and so provides a robust test of the more sophisticated theories that include these effects. At the energies investigated here it is also important to include target polarization and post-collisional interactions, which places further demands on the calculations. These effects reduce as the incident energy increases. The extended energy range of the DCS measurements presented here hence should allow their contributions to be considered systematically. 

The strength of the DWBA models described in~\cite{miller2015energy, al2009atomic} lies in the fact that different interaction processes can be turned `on' or `off' by replacing the distorted waves with plane waves. This allows different underlying scattering mechanisms to be considered. In the perpendicular plane these models suggest that peaks found at $\phi$ = 180$\degree$ are due to the momentum of the bound electron matching that of the incident electron, so that both electrons leave the interaction in opposite directions~\cite{Zhang1990}. Madison and co-workers~\cite{al2009atomic} further showed that peaks occurring at $\phi$ = 180$\degree$ will also have a contribution from ${triple}$ scattering, where the incident electron first scatters elastically into the perpendicular plane, followed by a binary collision with a bound electron. The electron then scatters elastically from the target to emerge at the mutual angle of 180$\degree$. Peaks found near $\phi$ = 90$\degree$ and 270$\degree$ are considered to arise from elastic scattering of the electron into the perpendicular plane, followed by a binary collision. Note that the DCS at angles $\phi$ $\leq$ 90$\degree$ are mirrored by the measurements for $\phi$ $\geq$ 90$\degree$, due to rotational symmetry around the incident electron beam direction.  

These semi-classical descriptions of the different collision mechanisms are attractive, as they provide an intuitive explanation of the processes that are involved. They have proven to be successful in describing ionization from lighter targets, however a full quantum calculation is needed to model the DCS for heavier atoms. The additional experimental data presented here are hence important to elucidate a better understanding of the scattering mechanisms that are involved.

These experiments proved to be challenging due to the very low coincidence count rates obtained from this target in the perpendicular plane. The rates varied from around 0.5 Hz to less than 0.01 Hz, depending on both the mutual angle and incident energy. The (e,~2e) spectrometer hence was operated under computer control, adopting optimization techniques to eliminate long-term drifts that can occur over the extended periods of time required to accumulate coincidence data. These control and optimization techniques have been described in depth previously~\cite{MurrayRSI, Patel_2020}.

The (e,~2e) spectrometer in Manchester uses an unselected-energy electron gun that produces an electron beam in the energy range from 5~eV to 300~eV with a width of around 0.6~eV. The electron beam enters the interaction region with zero beam angle and a pencil angle of around 2$\degree$. The electron analyzers use 3-element electrostatic zoom lenses that direct electrons emerging from the interaction into hemispherical energy selectors. The pass energies of the selectors are adjusted to control the overall resolution of the spectrometer and are typically set to match that of the gun. A gas jet directs atoms into the interaction region (not shown in Fig.~\ref{fig:e2ePerpendicularPlane}). The spectrometer is mounted inside a large $\mu$-metal enclosed chamber that is evacuated to a base pressure of $\sim~6$~x~10$^{-7}$~Torr. When the gas jet is injected the pressure rises to 2~x~10$^{-5}$~Torr. Further details of the spectrometer can be found in~\cite{Argon2022} and the references therein.

These new experiments were performed over a period of 10 months to accumulate the data presented here. Once the gas jet was set to deliver target atoms into the interaction region, the incident electron beam current was adjusted to produce scattered electron count rates in each analyzer that were between 5 and 10~kHz. The analyzers were then set to scan between $\phi$  = 70$\degree$ and 290$\degree$, this range being limited by their physical size. The lens voltages were optimized automatically at each angle to maximize the detected electron count rates. Coincidence counts were then accumulated for between 1,500 and 5,000~seconds, depending on the probability of detection of events at that energy and angle. The analyzers were then moved to a new angle and the process repeated. 

Once the detection plane had been mapped in one direction, the analyzers moved in the reverse direction. Up to 20 sweeps of the plane were conducted at a given energy. The measurements at each angle were then normalized to a fixed collection time and were averaged, the standard error in the mean providing the quoted uncertainty. These data were then re-normalized to unity at $\phi$ = 180$\degree$, as discussed above.

\begin{figure}[b]
\includegraphics[width=8.6cm]{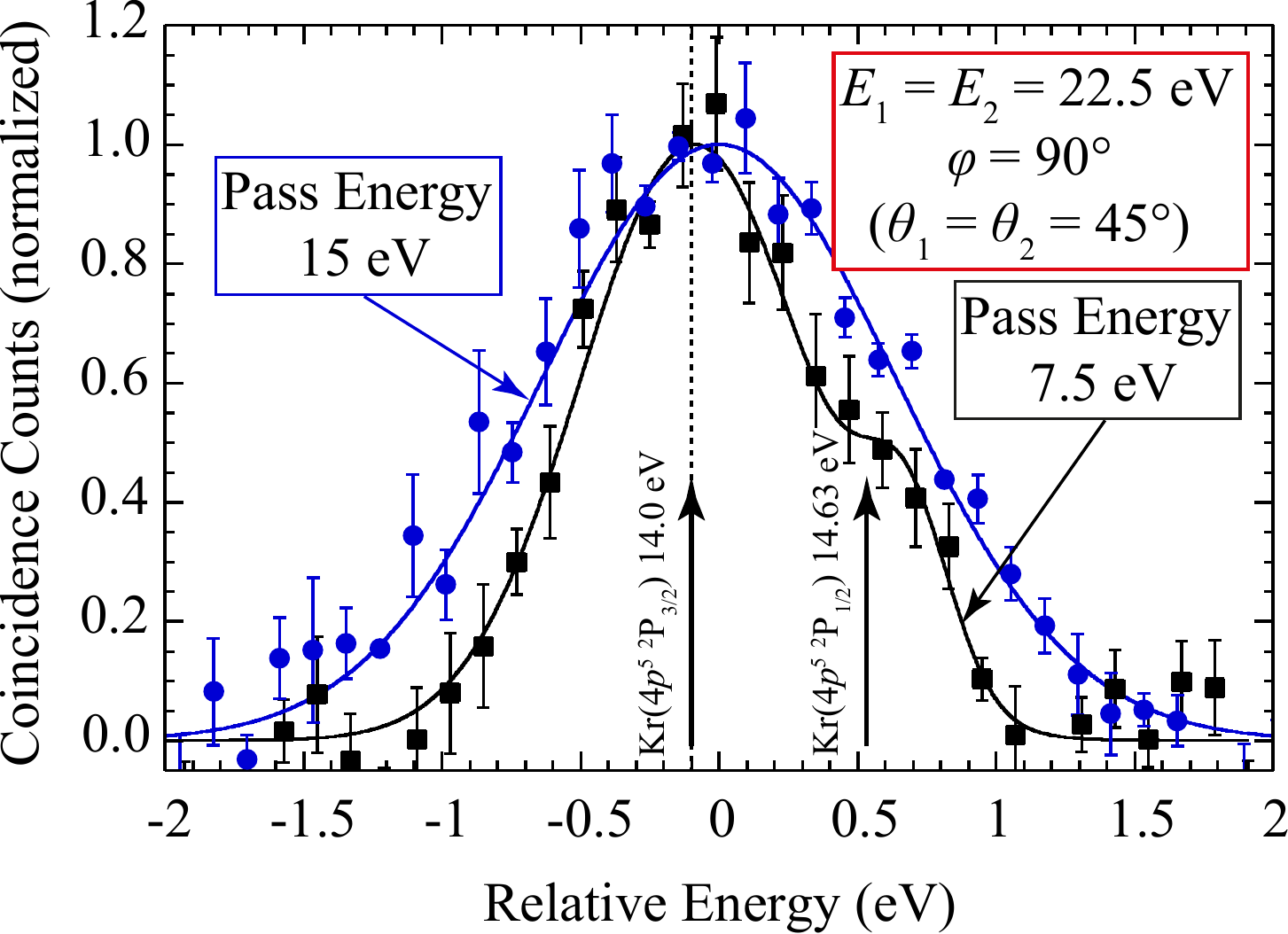}
\caption{\label{fig:EScan} The binding energy spectrum from Kr, where the outgoing energies were set to $E_1$ = $E_2$ = 22.5~eV and the mutual angle was $\phi$ = 90$\degree$ ($\theta_{1}$ = $\theta_{2}$ = 45$\degree$). The normalized data are shown for analyzer pass energies set to 15~eV and 7.5~eV respectively. At 7.5~eV the contribution from different core states can just be resolved.}
\end{figure}

All noble gas targets (apart from helium) have a complete outer valence shell comprised of 6 $p$-electrons. The ground state of krypton is the [Ar]3$d^{10}$4$s^2$4$p^6$~$^1$S$_{0}$ state, where [Ar] is the closed argon electron configuration. Removal of one of the outer electrons by electron-impact ionization hence leaves the core with five $p$-electrons, whose combined angular momentum is non-zero. This leads to two possible ionic core states, which are the 4$p^5$~$^2$P$_{3/2}$ and 4$p^5$~$^2$P$_{1/2}$ states. The $^2$P$_{3/2}$ core has the lowest binding energy of 14.0~eV, whereas the $^2$P$_{1/2}$ core has a binding energy $\sim0.63$~eV higher~\cite{NIST_2023}.

Since the resolution of the incident electron beam is around 0.6~eV it is possible to estimate the relative contribution from the different core states, by reducing the pass energy of the detectors so that the spectrometer resolution is dominated by that of the incident beam. Fig.~\ref{fig:EScan} shows the result of this study, where the outgoing electrons were detected at $\phi$ = 90$\degree$ ($\theta_{1}$ = $\theta_{2}$ = 45$\degree$) and were selected to have energies $E_1$ = $E_2$ = 22.5 eV.

The coincidence data in Fig.~\ref{fig:EScan} were obtained at pass energies of 7.5~eV and 15~eV respectively. The higher resolution data at 7.5~eV shows that the contribution from ionization to the 4$p^5$~$^2$P$_{1/2}$ core is around half that to the 4$p^5$~$^2$P$_{3/2}$ core under these kinematic conditions. Owing to the very low coincidence counting rates obtained in this geometry, it was however not practical to operate at the higher resolution and therefore all data presented in this paper were taken with a pass energy of 15~eV. Ionization to both channels hence needs to be considered when comparing these results to theory.

\section{DCS from 2 to 120~eV above the IP} \label{section:DCS2to120eV}

\begin{figure*}[ht]
\includegraphics[width=17.2cm]{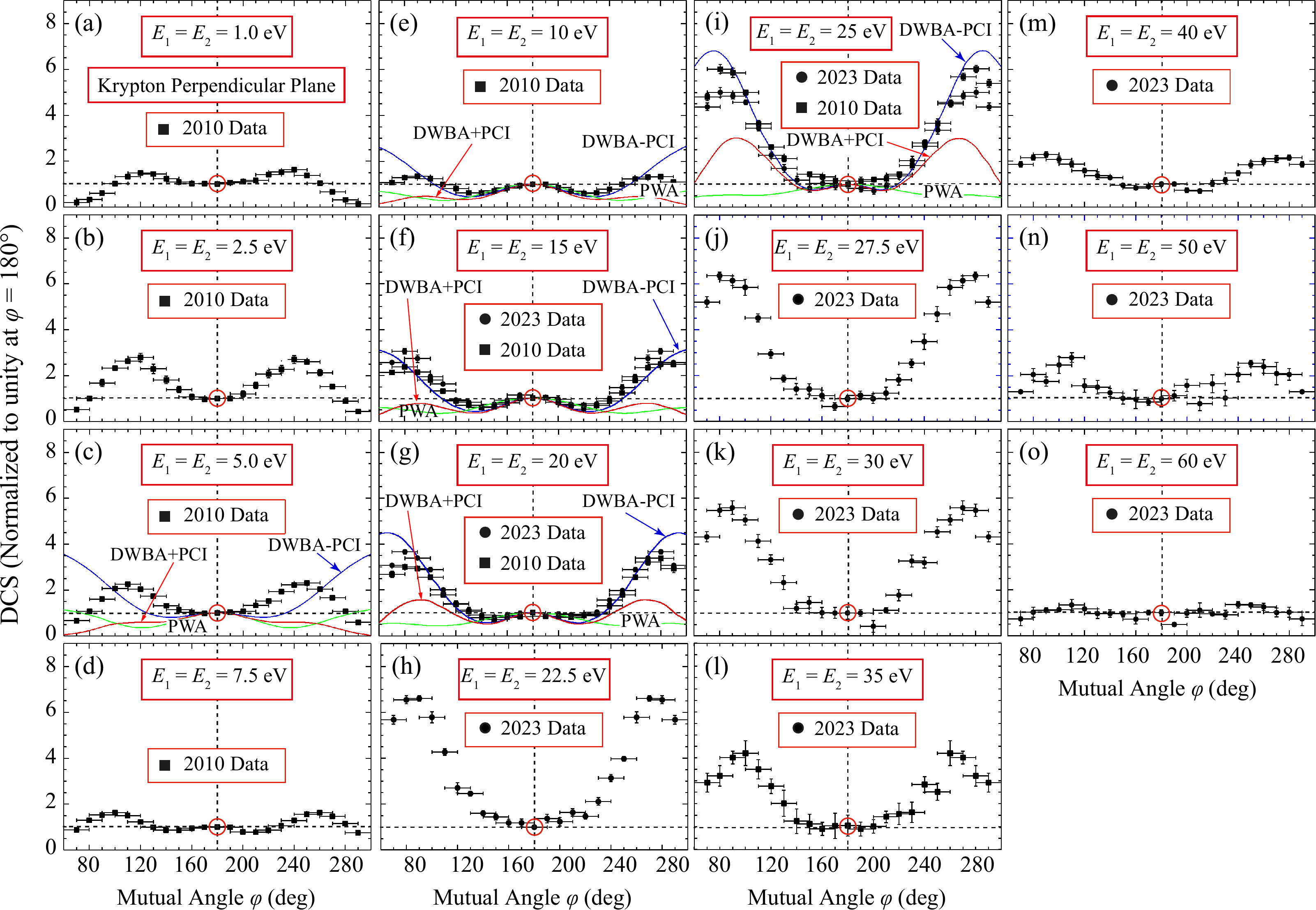}
\caption{\label{fig:2eVto120eV} Evolution of the DCS of krypton in the perpendicular plane from (a) 2~eV above the ionization potential to (o) 120~eV above the IP. The measurements are normalized to unity at $\phi$ = 180$\degree$. The data shown as black squares are from measurements taken in 2010 \cite{nixon2010low}. Data shown as filled black circles are new measurements. The calculations of \cite{miller2015energy} are also shown as solid curves, as discussed in section~\ref{section:DCS2to120eV}.}
\end{figure*}

The measured DCS for ionization of krypton are presented in Fig.~\ref{fig:2eVto120eV} for incident energies from 2~eV to 120~eV above the IP, using analyzer pass energies of 15~eV. The gun angle was set to $\psi$ = 90$\degree$ and the mutual angle $\phi$ was adjusted in steps of 10$\degree$. The results from \cite{nixon2010low} (2010 data) have been reproduced in Fig.~\ref{fig:2eVto120eV}, allowing comparison between the new measurements and previous data. 

This figure shows how the relative cross section evolves from a low incident energy 2~eV above the IP to a high energy 120~eV above the IP. The results from the calculations in~\cite{miller2015energy} are also reproduced where available, allowing comparison between theory and experiment. The DWBA calculation without PCI is the blue curve labelled DWBA-PCI. The calculation that includes PCI using the Gamow factor is the red curve labelled DWBA+PCI and the plane wave calculation is the green curve labelled PWA. Results from the calculations are included for incident energies of (c) 10~eV, (e) 20~eV, (f) 30~eV, (g) 40~eV and (i) 50~eV above the IP. 

The DWBA-PCI model produces the closest agreement with the data at the higher energies. Under these kinematic conditions the DCS must be zero at $\phi$ = 0$\degree$ and $\phi$ = 360$\degree$, due to post collisional interactions between outgoing electrons that have equal energy. The importance of this effect is clearly seen in the data. The DWBA-PCI model cannot however reproduce this condition, and so diverges from the data at low and high angles. At the lowest energy of 10~eV above the IP this model fails to emulate the data at all angles, as shown in Fig.~\ref{fig:2eVto120eV}(c).  

Inclusion of PCI using the Gamow factor in the DWBA+PCI calculation is clearly too strong an effect, since this does not reproduce the data well. The Ward Macek factor also produced too large a change compared to experiment and so for clarity is not reproduced here~\cite{miller2015energy}. Inclusion of PCI does however ensure that the cross section is zero at $\phi$ = 0$\degree$ and $\phi$ = 360$\degree$, as is required. The DWBA+PCI model predicts the position of the peaks reasonably well, however the magnitudes of the peaks around $\phi$ = 90$\degree$ and $\phi$ = 270$\degree$ are too small when normalized to the common point at $\phi$ = 180$\degree$.  

The plane wave approximation does not agree well with the data at any of the energies shown here. This demonstrates the importance of including the additional complexities of the interaction into the calculations as discussed above, particularly at lower energies. Since the PWA model is expected to become more accurate as the energy increases~\cite{weigold2012electron}, comparison with the new data at higher energies may show better agreement in the future.      

New results at an energy of 45~eV and ranging from 55~eV to 120~eV above the IP were obtained in this study, as shown in Fig. \ref{fig:2eVto120eV}(h) and panels (j) - (o). The complete set of data presented in Fig.~\ref{fig:2eVto120eV} hence details the evolution of the DCS in the perpendicular plane for Kr from near threshold to high energy. At the lowest observed energy with $E_1$ = $E_2$ = 1~eV the DCS has a relatively broad structure with small peaks occurring at around $\phi$ = 120$\degree$ and $\phi$ = 240$\degree$. This contrasts with measurements from the lighter noble gas targets He and Ne, which both show a large peak at $\phi$ = 180$\degree$ under similar kinematics~\cite{murray2000_neon,nixon2012mapping,Xenon_2022, murray_1997a,Bowring_1997,HawleyJones_1992,Selles_1987a,Selles_1987b}. Since the energy of the outgoing electrons are equal, the effects of PCI become increasingly important as the energy is lowered. This results in an enhancement of the back-to-back signal where the electrons emerge opposite each other (i.e. at $\phi$ = 180$\degree$). Near threshold, PCI is expected to dominate over all other collision processes, as described by Wannier~\cite{Wannier}. 

It is interesting to note that for Kr, Ar and Xe the DCS in this low energy regime does not feature a dominant peak at $\phi$ = 180$\degree$ as found for He and Ne and as predicted by the Wannier model. These targets all produce a `double peak' structure with a local $minimum$ at this angle, rather than a maximum. Since Ne also has $p$-electrons in the valence shell, this difference probably arises from the more complex electronic structure of these heavier targets. Further theoretical investigations in this energy regime are needed to explain these differences.   

As the incident energy increases beyond the threshold region the double-peak structure is enhanced until at 15~eV above the IP (Fig.~\ref{fig:2eVto120eV}(d)) a small central peak starts to emerge. Between 15~eV and 40~eV above the IP a triple peak structure is observed. When $E_1$ = $E_2$ = 10~eV all peaks have similar magnitudes. The side peaks then rapidly increase in magnitude compared to the central peak, until at 45~eV above the IP the central peak has disappeared. The remaining side peaks then slowly reduce in magnitude as the energy increases, until at 120~eV above the IP the DCS is largely uniform over a broad range of angles. This pattern is similar to that of the noble gases Ar and Xe at higher energies in this geometry \cite{Argon2022, Xenon_2022}. It is not clear why this flattening of the DCS occurs and so further calculations are required to explain these results.     

\subsection{Comparison between Ar and Kr peak ratios}

Given the differences between theory and experiment that are evident for Ar and Kr, it is instructive to consider how the ratio of the cross section measurements varies as the energy changes. Since the peak of the DCS occurs at around $\phi$ = 90$\degree$, the data at this angle (normalized to the common point at $\phi$ = 180$\degree$) can be compared as the energy changes. This ratio is sensitive to both the uncertainty in the 90$\degree$ measurement and that in the measurement at $\phi$ = 180$\degree$. The latter measurement has the larger uncertainty since the cross section at this angle is very small, with coincidence count rates often being less than 1 count in 100 seconds. Long accumulation time were hence required to reduce these uncertainties, as discussed above.  

\begin{figure}
\includegraphics[width=8.6cm]{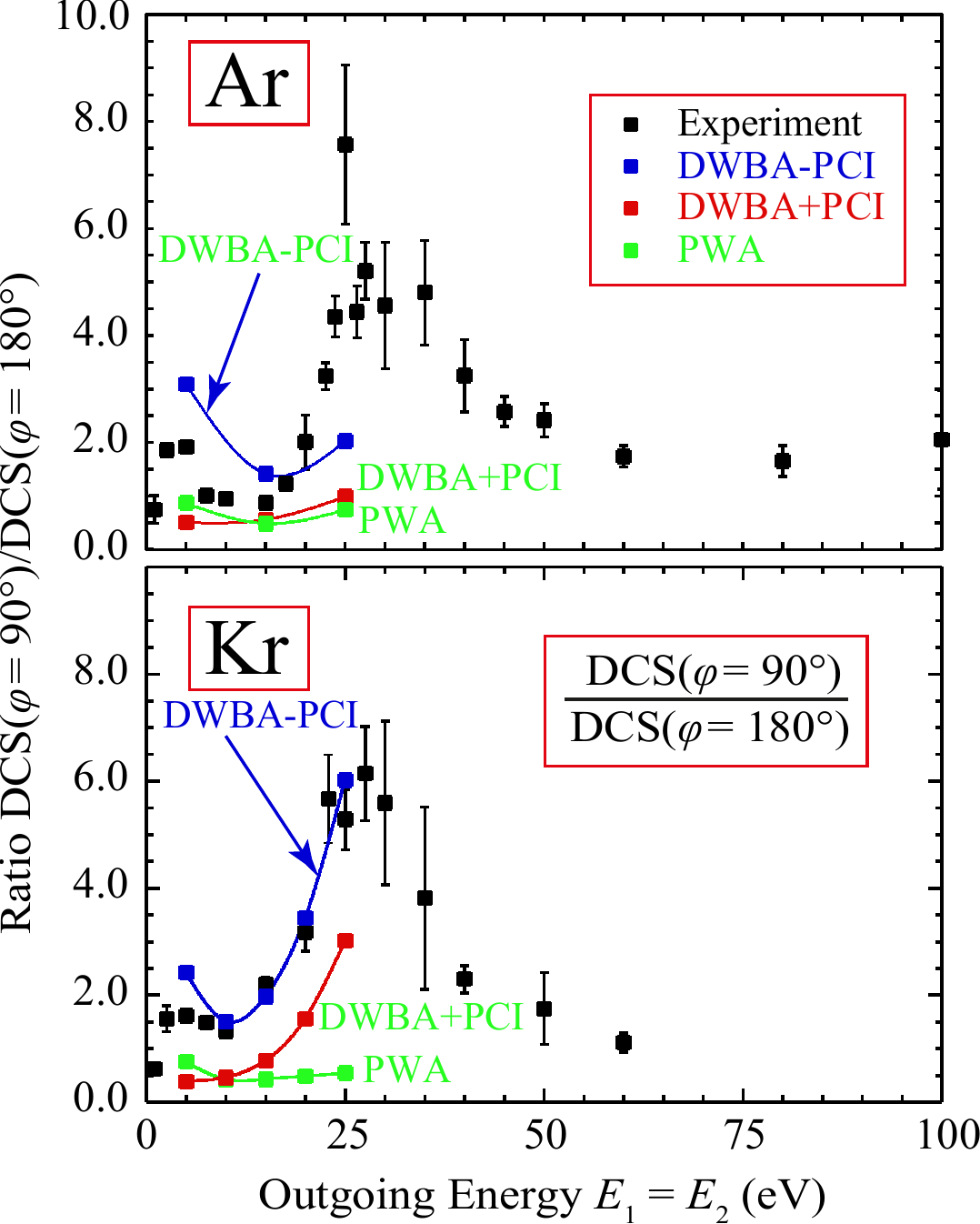}
\caption{\label{fig:ArKr_Ratios} Ratio measurements for the DCS at $\phi$~=~90$\degree$ to that at $\phi$~=~180$\degree$ for Ar and for Kr, taken from the measurements in Fig.~\ref{fig:2eVto120eV} and from~\cite{Argon2022}. The DWBA-PCI, DWBA+PCI and PWA calculations of Whelan and co-workers~\cite{miller2015energy} are also shown as solid curves for comparison.}
\end{figure}

Fig.~\ref{fig:ArKr_Ratios} shows the results of this analysis for both Ar and Kr over the range of energies where data were obtained. The results from the three different calculations of~\cite{miller2015energy} are also shown for comparison. The Ar measurements ranged from 5~eV to 200~eV above the IP for this target. The results from Kr were restricted between 2~eV to 120~eV above the IP but for comparison, the two are presented on the same energy scale.

It is interesting to note that the overall structure of these ratio measurements is similar between the targets. A small peak is seen near threshold in both cases, followed by a large rise which peaks at around 50~eV above the IP in each set of measurements. The ratio then slowly decreases, reaching a value of $\sim$1.9 for Ar at $E_1$ = $E_2$ = 100~eV and $\sim$1.1 for Kr at $E_1$ = $E_2$ = 60~eV. The uncertainties on these ratios are greatest at their peak values, due to the large uncertainty in the measurements at $\phi$~=~180$\degree$. There is however a similar structure observed in both sets of data over the same energy range.

The theoretical calculations from~\cite{miller2015energy} are only available at energies from 10~eV to 50~eV above the IP and these are shown as curves in Fig.~\ref{fig:ArKr_Ratios} to highlight their variations with energy. As expected, the PWA and DWBA+PCI calculations do not emulate the data well for either target. The DWBA-PCI calculation does not follow the data trend for Ar, however it closely matches the Kr results, apart from at the lowest energies. This agreement may be fortuitous, however these comparisons suggest that the differences between calculations should be investigated further.

\section{Summary and Conclusions}

In this paper the ionization cross sections for Kr have been presented over a range of energies from 2~eV to 120~eV above the ionization potential of the 4$^2$P$_{(1/2,3/2)}$ ion states. These states were unresolved in the experiments conducted here. A more detailed analysis of the binding energy spectrum (as shown in Fig.~\ref{fig:EScan}) indicates that both states contribute to the cross section and so need to be considered when comparisons to theory are made. All measurements were carried out in the perpendicular plane, the detected electrons being selected to have equal energies. The data are presented on a relative scale, with the DCS at $\phi$ = 180$\degree$ set to unity at each energy. 

The measurements are also compared to theoretical calculations from~\cite{miller2015energy} using both DWBA and PWA models. PCI was included via a Gamow factor, however this was found to produce too large a correction when compared to the data. Since PCI is important in this energy regime, it is essential for a more refined approach to PCI to be included in future models. The PWA model (which is successful at high incident energies) was found to be a poor approximation in this energy range for this target.

The DWBA-PCI model agrees closely with the ratio measurements between $\phi$~=~90$\degree$ and $\phi$~=~180$\degree$ for Kr. This model however does not agree with the Ar measurements, as shown in Fig.~\ref{fig:ArKr_Ratios}. This suggests these calculations should be revisited in order to understand this discrepancy.  

The DCS measurements at higher energies evolve into a broad flat structure, with the side lobes around $\phi$ = 90$\degree$ and 270$\degree$ reducing in magnitude as the energy increases. The broad, featureless cross section at the highest energy shown here has also been seen in other targets at similar energies, including Ar, Xe and CH$_{4}$~\cite{Harvey2019CH4}. The DCS in this region may hence be controlled by the kinematics, rather than by the structure of the target.  

It is hoped that the comprehensive survey of the DCS for Kr presented here will aid in the development and refinement of models of the ionization process, so that they can more accurately describe the interactions that are occurring in this important energy regime. 

\section{Acknowledgements}

We wish to thank the Engineering and Physical Sciences Research Council (EPSRC) for funding through grant EP/W003864/1 and EP/V027689/1. The data supporting the findings reported in this paper are openly available from the authors through the contact emails given above.

\bibliography{bibliographyKr5}

\end{document}